\documentclass[prl,showpacs,twocolumn,superscriptaddress,floatfix]{revtex4-1}
\usepackage{epsfig}
\usepackage{amsmath}
\usepackage{amssymb}
\usepackage{amsfonts}
\usepackage{mathptmx}
\usepackage{dcolumn}
\usepackage{eucal}
\usepackage{bm}
\usepackage{color}
\usepackage[colorlinks,linkcolor=blue,citecolor=blue]{hyperref}
\usepackage{epstopdf}
\usepackage{url}

\usepackage{dsfont}

\def\eps{\varepsilon}

\def\up{\uparrow}
\def\dw{\downarrow}

\def\be{\begin{equation}}
\def\ee{\end{equation}}
\def\bea{\begin{eqnarray}}
\def\eea{\end{eqnarray}}
\def\ba{\begin{array}{l l}}
\def\ea{\end{array}}

\begin{document}


\title{Zero magnetic-field orbital vortices in s-wave spin-singlet superconductors}

\author{Maria Teresa Mercaldo}
\affiliation{Dipartimento di Fisica ``E. R. Caianiello", Universit\`a di Salerno, IT-84084 Fisciano (SA), Italy}

\author{Carmine Ortix}
\affiliation{Dipartimento di Fisica ``E. R. Caianiello", Universit\`a di Salerno, IT-84084 Fisciano (SA), Italy}
\affiliation{Institute for Theoretical Physics, Center for Extreme Matter and Emergent Phenomena, Utrecht University, Princetonplein 5, 3584 CC Utrecht,  Netherlands}

\author{Francesco Giazotto}
\affiliation{NEST, Istituto Nanoscienze-CNR and Scuola Normale Superiore, Piazza San Silvestro 12, I-56127 Pisa, Italy}

\author{Mario Cuoco}
\affiliation{SPIN-CNR, IT-84084 Fisciano (SA), Italy, c/o Universit\`a di Salerno, IT-84084 Fisciano (SA), Italy}

\begin{abstract}
Breaking of time-reversal and 
point-group spatial symmetries can
have a profound impact on superconductivity. 
One of the most extraordinary effects, due to the application of a magnetic field, is represented by the Abrikosov vortices with charged supercurrents circulating around their cores.
Whether a similar phenomenon can be obtained by exploiting 
spatial symmetry breaking, e.g. 
through electric fields or mechanical strain, is a fundamentally relevant but not yet fully settled problem.
Here, we show that in two-dimensional spin-singlet superconductors with 
unusually low degree of spatial symmetry content, vortices with supercurrents carrying angular momentum around the core can form and be energetically stable. 
The vortex has zero net magnetic flux since it is
made up of counter-propagating Cooper pairs 
with opposite orbital moments. 
By solving self-consistently the Bogoliubov- de Gennes equations in real space, we demonstrate that the orbital vortex is stable and we unveil the spatial distribution of the superconducting order parameter around its core. The resulting amplitude has a characteristic pattern with a pronounced angular anisotropy that deviates from the profile of conventional magnetic vortices.
These hallmarks guide predictions and proposals for the experimental detection.
\end{abstract}
\maketitle


A vortex in a superfluid or superconductor is a point-like hole which around its core is marked by a phase gradient velocity field of the particles forming the bosonic or fermionic condensate. Since its original prediction \cite{Onsager1949} and investigation \cite{Feynman1955} in superfluid helium, vortices have been successfully observed in a broad range of systems including liquid helium\cite{Vinen1961,Bewley2008,Gomez2014}, ultracold atomic gases \cite{Hadzibabic2006}, photon fields \cite{Allen1992} and exciton-polariton superfluids \cite{Lagoudakis2008,Roumpos2011}. 
In superconductors, vortices carry quantized magnetic fluxes and typically arrange themselves in regular structures.
This prediction \cite{Abrikosov1957} has been firmly demonstrated by imaging and spectroscopic techniques \cite{Tonomura2001,Roditchev2015,Bonevich1994} in a large variety of so-called type-II superconductors, thus paving the way to a sparkling development of novel effects and phenomena with vortex quantum matter. 

While in superconductors the connection between vortices and magnetic field or time-reversal symmetry breaking is well settled, whether vortex structures can be induced by breaking spatial symmetries is an open and challenging problem. In this context, {structural and bulk} inversion symmetry breaking, leading to Rashba \cite{Rashba1960} and Dresselhaus \cite{Dresselhaus1955} spin-orbit coupling respectively, or lack of rotational or mirror symmetries, e.g. by 
built-in electric or strain fields, can play a relevant role. Non-centrosymmetric superconductors have been proposed to host anomalous Abrikosov vortices \cite{Hayashi2006,Garaud2020} but in the presence of applied magnetic fields, whose unconventional character can also arise from parity mixing of spin-singlet and spin-triplet pairs \cite{Gorkov2001,Frigeri2004}, magnetoelectric effects \cite{Levitov1985,Edelstein1995}, nonstandard Andreev states \cite{Vorontsov2008,Tanaka2009}, and topological phases \cite{Sato2009,Schnyder2010,Ying2017}.
As it concerns electrical manipulation, 
the recent observations \cite{DeSimoni2018} 
of critical supercurrent suppression in metallic superconductors by gating and of other anomalous effects \cite{Bours2020,Rocci2020,Paolucci2019,DeSimoni2019,Paolucci2018,Ritter2021,Alegria2021,Rocci2021,Golokolenov2021,Desimoni2021}, further underline a profound complexity, not yet fully uncovered, behind the coupling between superconductivity, point group symmetry breaking, 
and electric or strain fields.

In this Letter, we tackle this fundamental problem and demonstrate 
the formation in 2D superconductors with very low crystalline symmetry
of a vortex phase with zero net magnetic flux and spin-singlet pairs having non-vanishing orbital moments. This phenomenon relies on the direct coupling between the atomic orbital angular momentum ${\bf L}$ and the crystal wave-vector ${\bf k}$ via the
orbital-driven Rashba coupling \cite{Park2011,Park2012,Kim2013,Mercaldo2020}
due to crystalline potentials or externally applied electric fields that break spatial inversion.
We find that the orbital Rashba coupling ($\alpha_{{\text{OR}}}$) can induce and stabilize the nucleation of a vortex state made of counter-propagating Cooper pairs having total spin zero and opposite angular momentum 
associated with atomic orbital moments. 
Hallmarks of the orbital vortex state are represented by 
an inhomogeneous pattern of charge-neutral supercurrents with three-dimensional orbital moment textures, and a distinctive angular profile of the spatial amplitude of the order parameter around the vortex core. 
On the basis of symmetry constraints, electric field and/or strain turn out to be the most prospective means to generate and manipulate orbital vortices.   
 
\begin{figure}[bt]
\includegraphics[width=0.99\columnwidth]{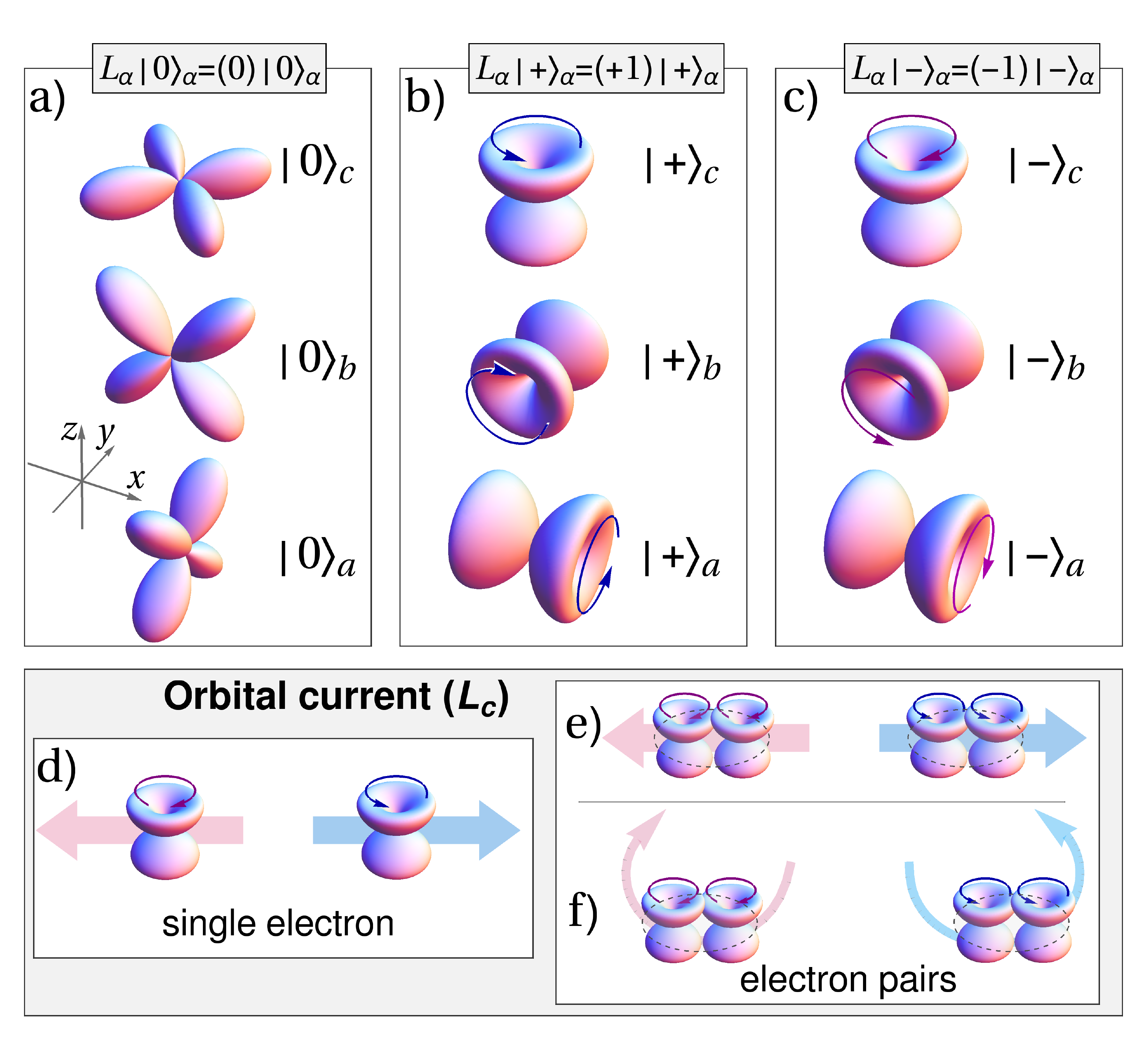} 
\protect\caption{(a) Illustration of the electron distribution associated with atomic orbitals having zero (a) and non-vanishing (b,c) projections of the angular momentum components $L_{\alpha}$ with $\alpha=(a,b,c)$ indicating the $(d_{xz},d_{yz},d_{xy})$ states. Sketch of $+1$ (b) and $-1$ (c) configurations with collinear orbital moments for various $\hat{L}$ components. Since the orbital density is the same for the $\pm$ states, we use a clockwise or anticlockwise arrow to indicate the opposite angular momentum along a given direction. Here, $|\pm\rangle_c=\frac{1}{\sqrt{2}}(\mp i |0\rangle_a+|0\rangle_b)$ and similar superpositions apply for $|\pm\rangle_{a}$,$|\pm\rangle_{b}$. Taking for example the $L_c$ projection, single electrons (d) or electron pairs (e) with opposite velocities and angular momentum lead to a current with non-zero orbital momentum. For an orbital vortex we have opposite sign in the circular velocity and orbital momentum (f).}
\label{f1}
\end{figure} 
 
{\it Model.} In order to address the role of point-group symmetries 
and orbital degrees of freedom in setting out a vortex phase, we consider a multi-orbital 2D electronic system that includes both mirror and rotation symmetry breaking interactions. A minimal description can be based on three bands arising from atomic orbitals spanning an $L=1$ angular momentum subspace, such as $p$ or $d$ orbitals. For convenience we refer to orbitals localized at the site of a square lattice (Fig. 1(a)) with $|0\rangle_{l}$  ($l=(a,b,c)$) representing Wannier states with zero angular momentum configurations, i.e. ${_\alpha}\langle 0|\hat{L}| 0\rangle_\alpha =0$ (Fig. \ref{f1}a). In this basis, the angular momentum components are given by $\hat L_k=i \eps_{klm}$, with $\eps_{klm}$ the Levi-Civita tensor. The breaking of inversion and of the horizontal mirror symmetry $\emph{M}_z$, {\it e.g.} due to structural inversion asymmetry and tunable by externally applied electric fields,
sets out an orbital Rashba interaction that couples the atomic angular momentum ${\bf L}$ with the crystal wave-vector ${\bf k}$. 
The residual vertical mirror symmetries ($\emph{M}_x$ and $\emph{M}_y$) as well as the $C_4$ rotation around the $\hat{z}$ axis can be broken for instance by local built-in elecric fields or inhomogeneous strains. This additional symmetry lowering
can be included by terms that involve the product of distinct orbital angular momentum components.
Hence, assuming a conventional s-wave spin-singlet pairing due to a local attraction, the Hamiltonian in real space can be expressed as \cite{Mercaldo2020,Bours2020,Mercaldo2021_a}:
\begin{eqnarray}
\mathcal{H}= \sum_{\langle i,j \rangle} \Psi^{\dagger}(i) \hat{H}(i,j) \Psi(j) \,,
\label{ham}
\end{eqnarray}
\noindent with 
\begin{eqnarray}
&& \hat{H}(i,j)= (\hat{t}_{i,j} - \mu \delta_{i,j}) \tau_z + \hat{\Delta} \tau_x \delta_{i,j} 
\end{eqnarray}
\noindent within the spinorial basis $\Psi^\dagger(i)=(\Psi_{\up}^\dagger(i), \Psi_{\dw}(i))$ assuming that
$\Psi_{\sigma}^\dagger(i)=(c^\dagger_{i,0_a,\sigma},c^\dagger_{i,0_b,\sigma},c^\dagger_{i,0_c,\sigma})$, $\tau_i$ are the Pauli matrices in the particle-hole space, $\delta_{i,j}$ is the Kronecker delta function. Here, $c_{r,0_\alpha\sigma}$ is the canonical electron annihilation operator at the position $r$, with spin $\sigma$ and zero orbital polarization components ($\alpha=(a,b,c)$).
The generalized hopping matrix is expressed as $\hat{t}_{i,j}=-t \hat{L}_a^2-i \alpha_{\text{OR}} \hat{L}_b+ \frac{\gamma}{2}(\hat{L}_c^2+\hat{L}_b^2-\hat{L}_a^2)+t_m [\{\hat{L}_a,\hat{L}_c\}+\{\hat{L}_b,\hat{L}_c\}]$ if the bond $<i,j>$ is along $x$, while $\hat{t}_{i,j}=-t \hat{L}_b^2-i \alpha_{\text{OR}} \hat{L}_a+ \frac{\gamma}{2}(\hat{L}_c^2+\hat{L}_a^2-\hat{L}_b^2)+t_m [\{\hat{L}_a,\hat{L}_c\}+\{\hat{L}_b,\hat{L}_c\}]$ for $y$ direction, with $\{\hat{A},\hat{B}\}$ indicating the anticommutator. $\alpha_{\text{OR}}$ is the strength of the orbital Rashba coupling whose structure in momentum space reads as $\hat{O}_{or}=\sin(k_x) \hat{L}_b - \sin(k_y) \hat{L}_a$ \cite{SupMat}. We also notice that upon a $C_4$ rotation $k_x \rightarrow k_y$, $k_y \rightarrow -k_x$ and the same applies to the angular momentum components, i.e. $\hat{L}_a \rightarrow \hat{L}_b$, $\hat{L}_b \rightarrow -\hat{L}_a$, $\hat{L}_c \rightarrow \hat{L}_c$. Furthermore, for the mirror symmetry $\emph{M}_y$ we have that $k_y \rightarrow -k_y$ ($k_x$ and $k_z$ stay unchanged) while the angular momentum is a pseudovector and hence the component perpendicular to the mirror plane does not transform, {\it i.e.} $\hat{L}_b \rightarrow \hat{L}_b$, while the other are inverted, i.e. $L_a \rightarrow -\hat{L}_a$ and $L_c \rightarrow -\hat{L}_c$.
On the basis of these symmetry transformations, we have that the orbital Rashba interaction term breaks only the horizontal mirror symmetry $\emph{M}_z$
while $t_m$ breaks the vertical mirrors and rotations. 
Note that $\gamma$ is an orbital dependent anisotropic term that is 
instead compatible with a $C_{4v}$ point group. 
Concerning the superconducting order parameter, we consider a conventional scenario with only local s-wave spin-singlet pairs with intra- and inter-orbital components. The general structure of $\hat{\Delta}$ in the orbital space can be expressed in the following form, $\hat{\Delta}=\frac{1}{2} \sum_{l,m} g_{\alpha \beta} \Delta_{\alpha \beta} (\hat{L}_{\alpha} \cdot \hat{L}_{\beta}  +\hat{L}_{\beta} \cdot \hat{L}_{\alpha}$) with $g_{\alpha \beta}$ being the orbital dependent pairing interaction and  $\Delta_{\alpha \beta}(r)= \langle c_{r,0_\alpha \uparrow} c_{r,0_\beta\downarrow} \rangle$ the related superconducting order parameters.

Here, $\langle ... \rangle$ stands for the ground-state expectation value. Due to the structure of $\hat{\Delta}$ we observe that the intra- and inter-orbital components of the order parameter (OP) have different symmetry properties. 
The intra-orbital terms $\Delta_{\alpha \alpha}$ 
are generally non-vanishing with the point-group symmetries that only pose constraints on their specific amplitudes.
On the other hand, since $\Delta_{\alpha\beta}$ is local and transforms as $(\hat{L}_{\alpha}\cdot \hat{L}_{\beta}+\hat{L}_{\beta}\cdot \hat{L}_{\alpha})$ 
then, it can change sign upon rotation and vertical mirror symmetry transformation, thus resulting into a vanishing amplitude. A lack of these symmetries is thus required to have an amplitude $\Delta_{\alpha\beta}$ different from zero. More details about the symmetry property of the superconducting order parameters are reported in the supplemental material \cite{SupMat}. 
Such low symmetry content
can be encountered at the surface or interface due to, {\it e.g.},
atomic termination profiles, strain or external electric field gradients. 
Alternatively, the point group symmetries can be intrinsically broken by an unusually low crystalline symmetry arrangement.

Now, we show that, if these order parameters have a non-trivial spatial dependence, akin to 
pair density waves with specific relations among their amplitudes and relative phases,
a vortex state with electron pairs having non-vanishing and collinear orbital moments can be obtained. This outcome is directly accessible by firstly observing that for any component of the orbital angular momentum one can construct pairs with $\pm2$ total projection. Indeed, for electron pairs with parallel orbital moments oriented along $\alpha$ we have that
$\eta_{+,\alpha}(r)=\langle c_{r,+_{\alpha} \uparrow} c_{r,+_{\alpha} \downarrow} \rangle$ and
$\eta_{-,\alpha}(r) =\langle c_{r,-_\alpha,\uparrow} c_{r,-_\alpha,\downarrow} \rangle  \,$. 
Then, we notice that from the structure of the single particle states having $\pm 1$ orbital moment, $\eta_{\pm}$ is expressed as a combination of intra- and inter-orbital pairing amplitudes. For instance, $\eta_{\pm,c}$ is given by
\begin{eqnarray}
\eta_{\pm,c}(r)=\left[ \frac{1}{2} \left(-\Delta_{aa}(r) + \Delta_{bb}(r)\right) \pm i \Delta_{ab}(r) \right]\,, \label{triplet}
\end{eqnarray}
\noindent 
and similar expressions 
can be obtained for the other components $\eta_{\pm,a}$ and $\eta_{\pm,b}$ \cite{SupMat}.
Hence, to describe a vortex phase with electron pairs having equal orbital moments (i.e. orbitally polarized), the order parameter has to be expressed as $\eta_{\pm}(r)=f(r) \exp[\pm i \theta(r)]$, whose non-trivial phase winding has opposite sign for the corresponding $\pm$ orbital angular momentum, while the spatial dependent amplitude is encoded in $f(r)$. 
Here, the phase winding is an orbital coherent effect which is evident in the structure of $\eta_{\pm}$ through its real and imaginary components, with $\eta_{\pm}$ being time-reversal partners. 
With the above elements, the vortex state can be directly obtained by considering the angular variable $\theta(r)$ for the vector ${\bf (r-r_0)}=(x-x_0,y-y_0)$ in the $x$-$y$ plane with respect to the center ${\bf r}_0$, and let the intra- and inter-orbital order parameters have a pair density modulation in real space such as 
$\Delta_{\alpha\alpha}(r)-\Delta_{\beta\beta}(r)=2 f(r) \cos \theta(r)$
and $\Delta_{\alpha\beta}(r)=f(r) \sin \theta(r)$. 

Next, we
present one specific hallmark of the vortex state, which is directly related to its orbital moment character and the presence of orbital currents. 
\begin{figure}[bt]
\includegraphics[width=0.99\columnwidth]{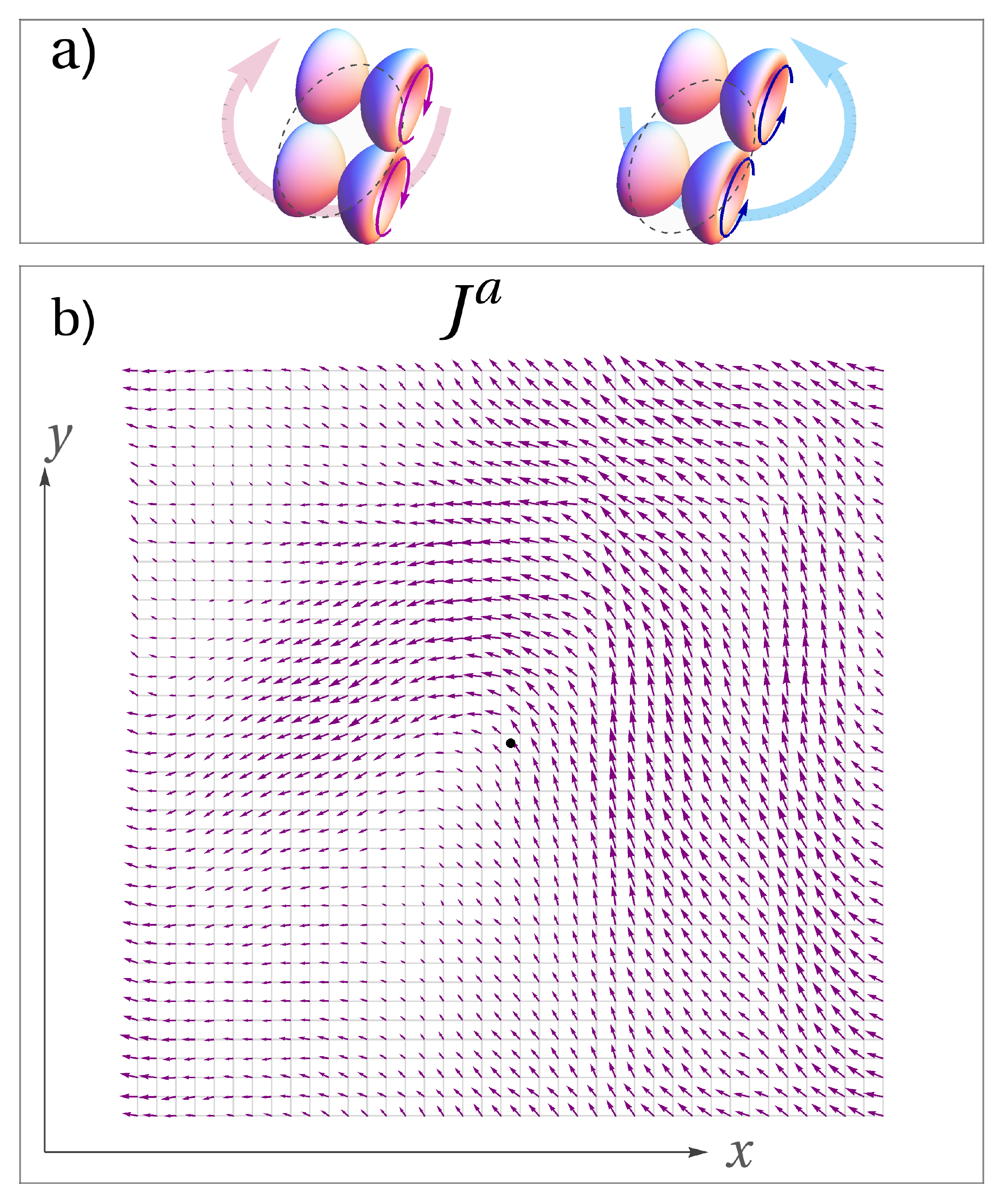} 
\protect\caption{(a) Illustration of electron pairs with non-zero orbital moment having reversed sign of the angular momentum projections and opposite velocity thus leading to a net orbital supercurrent without charge flow. (b) Spatial distribution of the orbital supercurrent around the vortex core (black dot) for the current $J^a$ carrying orbital moment oriented along the $x$ direction. The arrow indicates the direction of the orbital current on each bond, and the length its amplitude. The vortex state is marked by a non-trivial phase winding for $\eta_{\pm,b}(r)$ and $\eta_{\pm,a}(r)$ order parameters. The order parameter $\eta_{\pm,c}(r)$ with $c$-oriented orbital moment has zero amplitude. 
The simulation refers to a system size $N_x \times N_y$ with $N_x=N_y=40$ while the following parameters of the 2D model Hamiltonian have been used: $\alpha_{\text{OR}}=1.0 t$,$\gamma=0.1 t$,$t_m=0.2 t$,$g\Delta_0=0.1 t$, $\mu=-0.4 t$ and for the order parameters $f_0=0.3$,$f_1=f_2=0.5$, and $g_{\alpha\beta}=g=2t$.}
\label{f2}
\end{figure} 
As for the spin-current \cite{Nikolic2006}, one can introduce orbital-current operators on a bond $\langle i,j \rangle$ for the various orbital moment components. We focus on the leading term for the transfer of orbital angular momentum, i.e. $\hat{J}^{\alpha}_{ij}=i(\,c^{\dagger}_{i,0_m \sigma} \hat{L}_{\alpha}^{m,m'} c_{j,0_{m'}\sigma} -\text{h.c.})$ \cite{SupMat}.
We then evaluate the distribution of the orbital current in the superconducting state around the vortex core. 
We have selected a representative 
set for 
the local superconducting order parameters 
assuming that $\Delta_{aa}=\Delta_{bb}\neq \Delta_{cc}$ and $\Delta_{ac}=\Delta_{bc}$, while $\Delta_{ab}=0$. 
For convenience, the parameterization in real space for $\hat{\Delta}$ at a given position, $(x_i,y_i)$ is expressed as $[\Delta_{aa(bb)}(x_i,y_i)-\Delta_{cc}(x_i,y_i)]=f_1 \Delta_0 \cos(\theta(x_i,y_i))$, $\Delta_{cc}(x_i,y_i)=f_0 \Delta_0$ and $\Delta_{ac(bc)}= f_2 \Delta_0 \sin(\theta(x_i,y_i))$, where $\Delta_0$ sets the overall scale and $f_0$, $f_1$, $f_2$ are coefficients that take into account the possible orbital anisotropy of the superconducting state. Variations in the $f$ values do not affect the qualitative outcome of the orbital current profile.

In the absence of vortices, we have a uniform equilibrium orbital supercurrent with in-plane directional character, i.e. it substantially transfers Cooper pairs with angular momentum parallel to $b(c)$ when propagating along $x(y)$, respectively. The current with an out-of-plane orbital moment is vanishing for the uniform solution. On the contrary, the pattern of orbital supercurrents exhibits a characteristic winding of the angular momentum flow due to the presence of the vortex (Figs. \ref{f2}b). The vorticity is regular and of the same type for $J^a$ and $J^b$ while it has a sort of more turbulent aspect for $J^c$ \cite{SupMat}. 
The circulation of orbital moments are fingerprints of the superconducting orbital vortex.  
We notice that the flow of the orbital supercurrent does not show a complete winding around the vortex core. This is due to the fact that in the ground state there are configurations which are reminiscent of the uniform pattern of orbital currents in the normal state. Then, these contributions tend to give a cancellation in some regions amplifying the amplitude asymmetry around the core (Figs. \ref{f2}b).

\begin{figure}[bt]
\includegraphics[width=0.9\columnwidth]{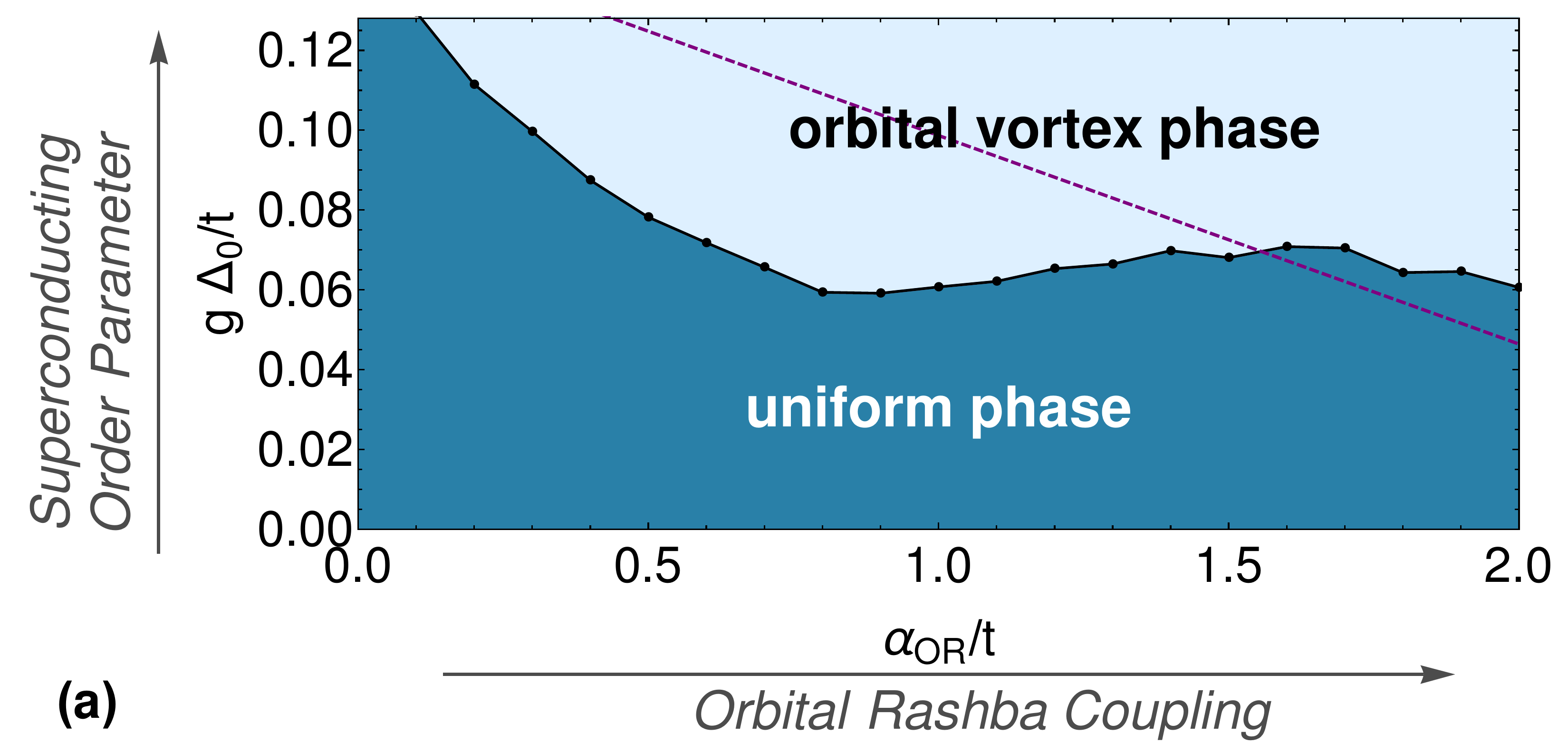} 
\includegraphics[width=0.9\columnwidth]{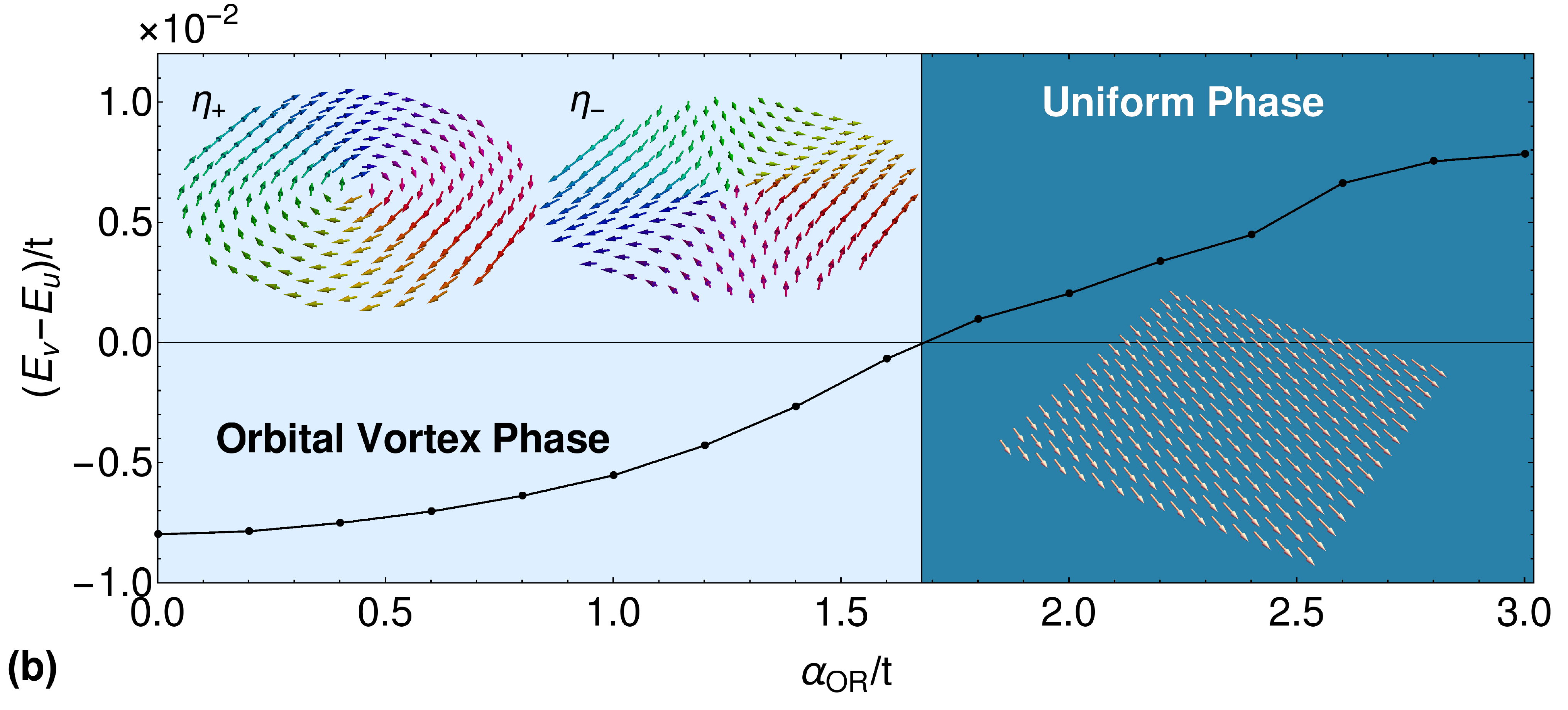} 
\includegraphics[width=0.9\columnwidth]{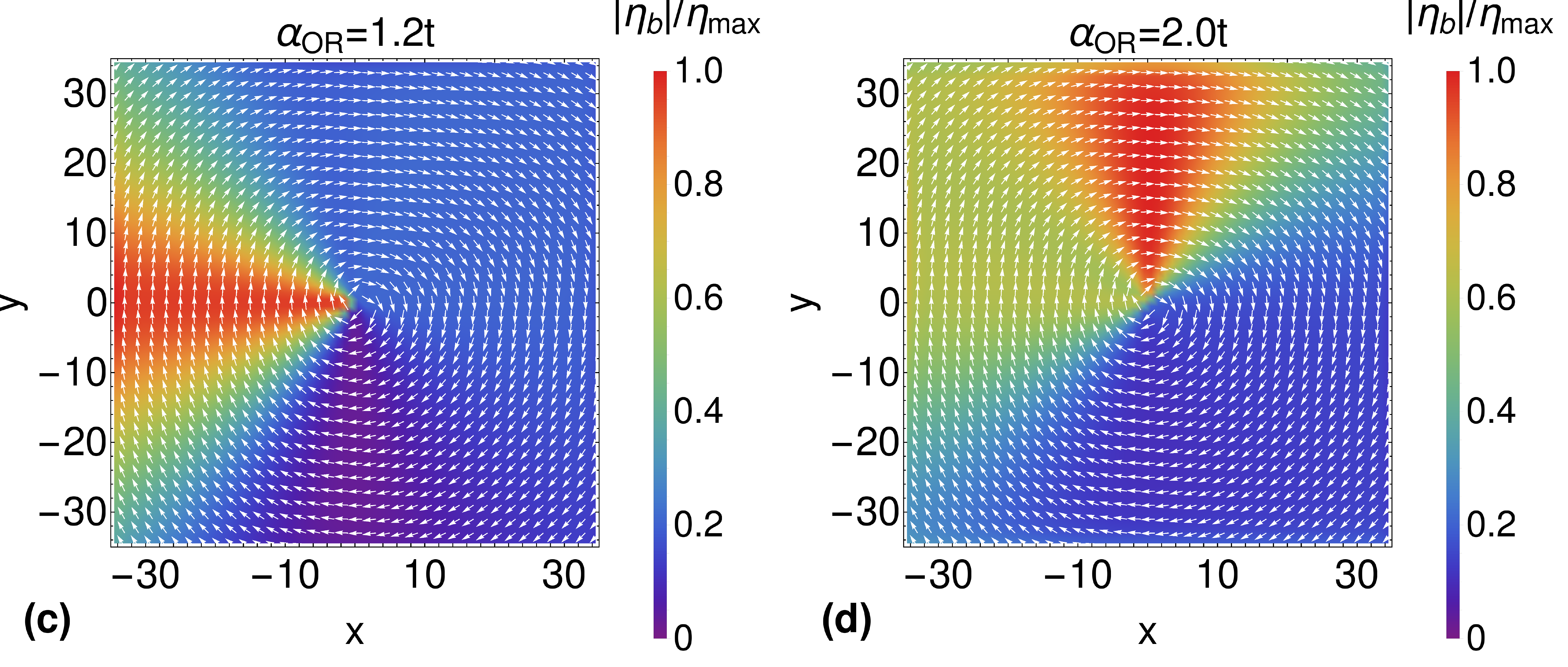} 
\protect\caption{Phase diagram with uniform and orbital vortex configurations as a function of the energy scale ($g\Delta_0$), setting the amplitudes of superconducting order parameters $\Delta_{\alpha\beta}$, and the orbital Rashba interaction ($\alpha_{\text{OR}}$). 
The simulation refers to a system size $N_x \times N_y$ with $N_x=N_y=70$, while the model parameters are the same as in Fig.\ref{f2}. 
(b) Energy difference between the vortex solution ($E_v$) and the uniform one ($E_u$) as a function of the orbital Rashba coupling $\alpha_{\text{OR}}$.
Both solutions have been obtained evaluating the SC OP components 
$\Delta_{\alpha \beta}$ with an iterative self-consistent procedure in real space for a system of size $N_x=N_y=20$. In the self-consistent analysis the  components of the SC OP decrease by increasing $\alpha_{\text{OR}}$ (see \cite{SupMat}). This trajectory is approximately given by the purple dashed line drawn in (a). In the insets we have a graphical representation of the order parameters ($\eta_{\pm}$) with opposite winding in the vortex, assuming that the vector components are the corresponding real and imaginary parts. 
(c-d) Density plot of the amplitude $\eta_b$ for two values of $\alpha_{\text{OR}}$, obtained self-consistently for a system of size 
$70 \times 70$. $|\eta_b|$ shows a non-uniform spatial behavior, with a distribution which changes varying $\alpha_{\text{OR}}$. 
The white arrows give a representation of the winding of $\eta_{+,b}$, where the vector components are the real and imaginary part of $\eta_{+,b}/|\eta_{b}|$. For clarity of visualization the arrows are not drawn at all the sites. The spatial profile for $\eta_{\pm,a}$ is the same of $\eta_{\pm,b}$ while $\eta_{\pm,c}$ has a negligible amplitude.
The parameters are $\gamma=0.1 t$,$t_m=0.2 t$, $\mu=-0.4 t$, $g_{ab}=0$, and for the other bands $g_{\alpha\beta}=2t$. As supplemental material we provide movies about the evolution of $\eta_{\pm,b}$ by varying $\alpha_{\text{OR}}$.}
\label{f3}
\end{figure}

Let us move to the energetics of the orbital vortex state. The strategy here is to firstly analyze the energy competition between the vortex and uniform states by assuming a homogeneous profile for the order parameters. Then, we determine the vortex configuration solving 
self-consistently the equations in real space for the local amplitudes
of all the order parameters.
In Fig. \ref{f3}a we present the phase diagram obtained by comparing the free energy of the uniform phase with that of an orbital vortex state having uniform amplitude of the order parameter. 
We set an amplitude $\Delta_0$ to fix the overall scale for intra- and inter-orbital order parameters. Since the coherence length decreases when $\Delta_0$ grows one can effectively span different regimes of $\xi_S/L$, as for BCS superconductors $\Delta_0 \sim 1/\xi_S$ (in the performed simulation for $g \Delta_0/t \sim 0.05$ we have $\xi_S/L \sim 1$). The simulation is performed for a system size $N_x \times N_y$, where $N_x=N_y=L$ and $L=70$ in units of the interatomic distance. We consider the orbital Rashba coupling as the main varying parameter and assume a given amplitude of the coupling ($t_m=0.2 t)$ that breaks mirrors and rotational symmetries. 
The outcome indicates that there is a boundary line of critical $\alpha_{\text{OR}}$ separating the region of phase space where the vortex is energetically more stable than the uniform state. The $\alpha_{\text{OR}}$ coupling can lead to a transition into the vortex phase with uniform amplitude. 
For sufficiently low values of the superconducting interaction and superconducting gap, corresponding to $\xi_S/L \sim 1$, we find that the orbital Rashba interaction becomes substantially ineffective in sustaining the orbital vortex phase. This observation qualitatively implies that in superconductors with low critical temperature the nucleation of orbital vortices might be less favorable.

To demonstrate the stability of the vortex state and to corroborate this trend obtained by making an ansatz on the amplitude of the superconducting order parameters, we performed a full self-consistent analysis of the spatial dependent order parameters. In Fig. \ref{f3}b we compare the energy of the solutions obtained by self-consistently solving the gap equations at each lattice site for the $\eta_{\alpha}$ assuming either a vortex phase winding or a spatially uniform phase for the superconducting order parameters. The presented simulation is for a system size $N_x=N_y=20$. However, we have verified for few values of $\alpha_{\text{OR}}$ that for larger size, up to $N_x=N_y=70$, the stability is preserved \cite{SupMat}. 
We observe that above a critical amplitude of $\alpha_{\text{OR}}$ the vortex state becomes unstable. This behavior can be understood by noticing that $\alpha_{\text{OR}}$ is suppressing the amplitude of the SC-OPs and thus brings the system into a regime where the orbital Rashba coupling is unable to stabilize the vortex state. Such self-consistent trajectory of the OP is schematically sketched in Fig. \ref{f3}a. 
The analysis of the amplitude of $\eta$ in the vortex state indicates a peculiar angular anisotropy (Figs. \ref{f3}c,d) with a modulation of intensity from small to large values as one moves around the center of the vortex. The pattern of $\eta$ is consistent with mirror and rotational symmetry breaking. The position of maximal amplitude of the order parameter is typically pinned to the high symmetry lattice directions thus having possible implications concerning the formation of domains. 

Let us finally discuss how to detect this type of vortex and few direct consequences of their presence.
The orbital vortex has zero net magnetic flux.
However
since neutral currents of magnetic dipoles can produce electric fields \cite{Sun2004}, we predict that the vortex state of orbital moments can exhibit a weak electric field, at a given distance $R$ from the orbital current sources, according to the following relation ${\bf E}=\int d^3 r \frac{\mu_0}{4 \pi} {\bf J}_{\alpha} \times \frac{1}{R^3} [\hat{n}_{\alpha} - \frac{ 3 {\bf R} ({\bf R}\cdot \hat{n}_{\alpha})}{R^2}]$ \cite{Sun2004}, with ${\bf J}_{\alpha}$ and $\hat{n}_{\alpha}$ the orbital current and orbital moment associated with the $\alpha$ orientation, respectively. This electric field would be typically screened on a distance of the order of the Thomas-Fermi length. 
On the basis of the patterns in Fig. \ref{f2}, especially outward to surface or along directions of poor screening inside the material, around the core of the vortex there would be the most prominent features of an electric field distribution with in-plane and out-of-plane components. 

Another path to unveil the presence of vortices with orbital moments is to make use of weak perturbations that break time-reversal symmetry
such as external currents or small magnetic fields. In this case, the lack of time reversal symmetry would lead to uncompensating effects in the vortex orbital flow and weak orbital moments of the Cooper pairs that would result into anomalous magnetic flux response. The flux is not quantized and can be directly probed by means of SQUID or electron microscopy techniques. 
Furthermore, to detect the presence of orbital vortices one could also employ scanning tunneling microscopy (STM). On the basis of the achieved self-consistent patterns (Figs. \ref{f3}c,d), we expect that a variation of the orbital Rashba coupling has an impact on the spatial distribution of the superconducting order parameter nearby the vortex core. This implies that spatially resolved STM measurements would detect an enhanced population of in-gap quasiparticles inside the vortex with different characteristic lengths even in the absence of external magnetic field. 

If the crystalline environment does not provide symmetry breaking interactions that are sufficiently strong to allow for the nucleation of the orbital vortex, we expect that the application of external electric fields or strains can be the most impactful means
to yield the orbital vortex phase.  
Hence, we foresee the achievement of an electric- or strain-driven dual of the magnetic vortex phase that can open a wide avenue of explorations of quantum vortex phases in ultrathin superconductors. 

\begin{acknowledgments}
M.C., M.T.M and F.G. acknowledge support by the EU’s Horizon 2020 research and
innovation program under Grant Agreement nr. 964398 (SUPERGATE). 
C.O. acknowledges support from a VIDI grant (Project 680-47-543) financed by the Netherlands Organization for Scientific Research (NWO). F.G. acknowledges the European Research Council under Grant Agreement No. 899315-TERASEC, and  the  EU’s  Horizon 2020 research and innovation program under Grant Agreement No. 800923 (SUPERTED) for partial financial support.
We acknowledge valuable discussions with A. Caviglia and A. di Bernardo.
\end{acknowledgments}

\end{document}



\title{Supplemental Material for ``Zero magnetic-field orbital vortices in s-wave spin-singlet superconductors''}

\author{Maria Teresa Mercaldo}
\affiliation{Dipartimento di Fisica ``E. R. Caianiello", Universit\`a di Salerno, IT-84084 Fisciano (SA), Italy}

\author{Carmine Ortix}
\affiliation{Dipartimento di Fisica ``E. R. Caianiello", Universit\`a di Salerno, IT-84084 Fisciano (SA), Italy}
\affiliation{Institute for Theoretical Physics, Center for Extreme Matter and Emergent Phenomena, Utrecht University, Princetonplein 5, 3584 CC Utrecht,  Netherlands}

\author{Francesco Giazotto}
\affiliation{NEST, Istituto Nanoscienze-CNR and Scuola Normale Superiore, Piazza San Silvestro 12, I-56127 Pisa, Italy}

\author{Mario Cuoco}
\affiliation{SPIN-CNR, IT-84084 Fisciano (SA), Italy, c/o Universit\`a di Salerno, IT-84084 Fisciano (SA), Italy}


\begin{abstract}
\noindent 
In the Supplemental Material we provide details about the model and the vortex solutions. We present the structure of the orbital Rashba interaction and the orbital current. We discuss the symmetry aspects of the intra- and inter-orbital superconducting order parameters with respect to the point-group spatial transformations. We show the phase diagram based on vortex configurations with and without self-consistent solutions. We present the dependence of the self-consistent solution for the superconducting order parameters as a function of the orbital Rashba coupling. Finally, we analyze the energetics of the vortex state evaluated self-consistently by varying the size of the system.
\end{abstract}
\maketitle

\section{Orbital Rashba coupling and orbital currents}

{\it Orbital Rashba coupling.} In a two-dimensional system the lack 
mirror symmetry due to the $z\rightarrow -z$ transformation, and consequently of inversion,
can be generally accounted by a potential $V(z)$ that does not have a definite parity. In a multi-orbital configuration, the antisymmetric part of the potential $V(z)$ yields an orbital mixing that is the orbital analogue of the spin Rashba interaction. Assuming for instance a square lattice, the coupling is substantially due to the fact that, for two sites at position $R_i$ and $R_j$, the matrix elements of $V(z)$ between atomic orbitals that are linked via the $\hat{L}_b$ or $\hat{L}_a$ components of the orbital angular momentum are odd as a function of $R_i-R_j$ \cite{Mercaldo2020,Bours2020}. Indeed, this type of orbital hybridization on a square lattice leads to an orbital Rashba interaction in momentum space of the form $\alpha_{\text{OR}} \,\hat{g}(k)\cdot \hat{L}$, with $\hat{g}(k)=\{\sin k_y,-\sin k_x,0\}$ being the inversion asymmetric vector.

{\it Orbital current operator.} To introduce the orbital-current operator one can follow the same approach that is implemented for the spin-Rashba coupling \cite{Nikolic2006}. 
The orbital angular momentum density in the lattice models is given by the local orbital operator $\hat{L}_n=\{\hat{L}_{a,n},\hat{L}_{b,n},\hat{L}_{c,n} \}$ at a site indicated by the position vector $n$ as
\begin{eqnarray}
\hat{L}_{\alpha,n}=\hbar \sum_{i,j} c^{\dagger}_{n,0_i} \hat{L}_{\alpha}^{ij} c_{n,0_j}
\end{eqnarray}
\noindent where for clarity we have dropped the reference to the spin index and $\alpha=\{a,b,c\}$ refers to the orientation of the orbital moment. Here, the basis in the orbital space is made of configurations with zero projection of the angular momentum.
Without loss of generality, one can assume that the single particle Hamiltonian is expressed as $\hat{\cal{H}}=\sum_{m,m'} c^{\dagger}_{m,0_\alpha} \hat{t}_{m,m'} c_{m',0_\beta}+\text{h.c.}$ with $\hat{t}_{m,m'}$ being a function of $\hat{L}_{\alpha}$ and $\hat{L}^2_{\alpha}$ for a generic bond identified by $(m,m')$ with $m'=m+e_i$, $i=x,y$ and $e_i$ being the unit vectors basis. 
The Heisenberg equation of motion for each component of the local orbital operator  
\begin{eqnarray}
\frac{d \hat{L}}{dt}=-i \hbar [\hat{L},\hat{H}]
\end{eqnarray}
\noindent can be written in the following form 
\begin{eqnarray}
\frac{d \hat{L}_{\alpha,m}}{dt}+ \sum_{k} \left( \hat{P}^{L_{\alpha}}_{m,m+e_k}-\hat{P}^{L_{\alpha}}_{m,m-e_k}  \right)=\hat{F}_{m}^{L_{\alpha}}
\end{eqnarray} 
where $\hat{P}^{L_{\alpha}}_{m,m'}$ is the bond operator given by 
\begin{eqnarray}
\hat{P}^{L_{\alpha}}_{m,m'}=\sum_{p,q} \frac{1}{4 i} \left[ c^{\dagger}_{m,0_p} \{ \hat{L}_{\alpha}, \hat{t}_{m,m'} \}^{pq}  c_{m',0_q} \right ]
\end{eqnarray}
and $\hat{F}_{m}^{L_{\alpha}}$ is the orbital source operator. The structure of the equation of motion of the orbital operator recalls that one for the continuity equations for the spin density.
By considering the operator $\hat{P}^{L_{\alpha}}_{m,m'}$ and the source term $\hat{F}_{m}^{L_{\alpha}}$ we have that the character of the orbital current flow can be captured by evaluating all the terms of the tensor $\hat{J}^{\alpha}_{mm'}=i \sum_{p,q} (\,c^{\dagger}_{m,0_p \sigma} \hat{L}_{\alpha}^{p,q} c_{m',0_{q}\sigma} -\text{h.c.})$, as we have done for the orbital vortex state.

\begin{figure*}[bt]
\includegraphics[width=0.49\textwidth]{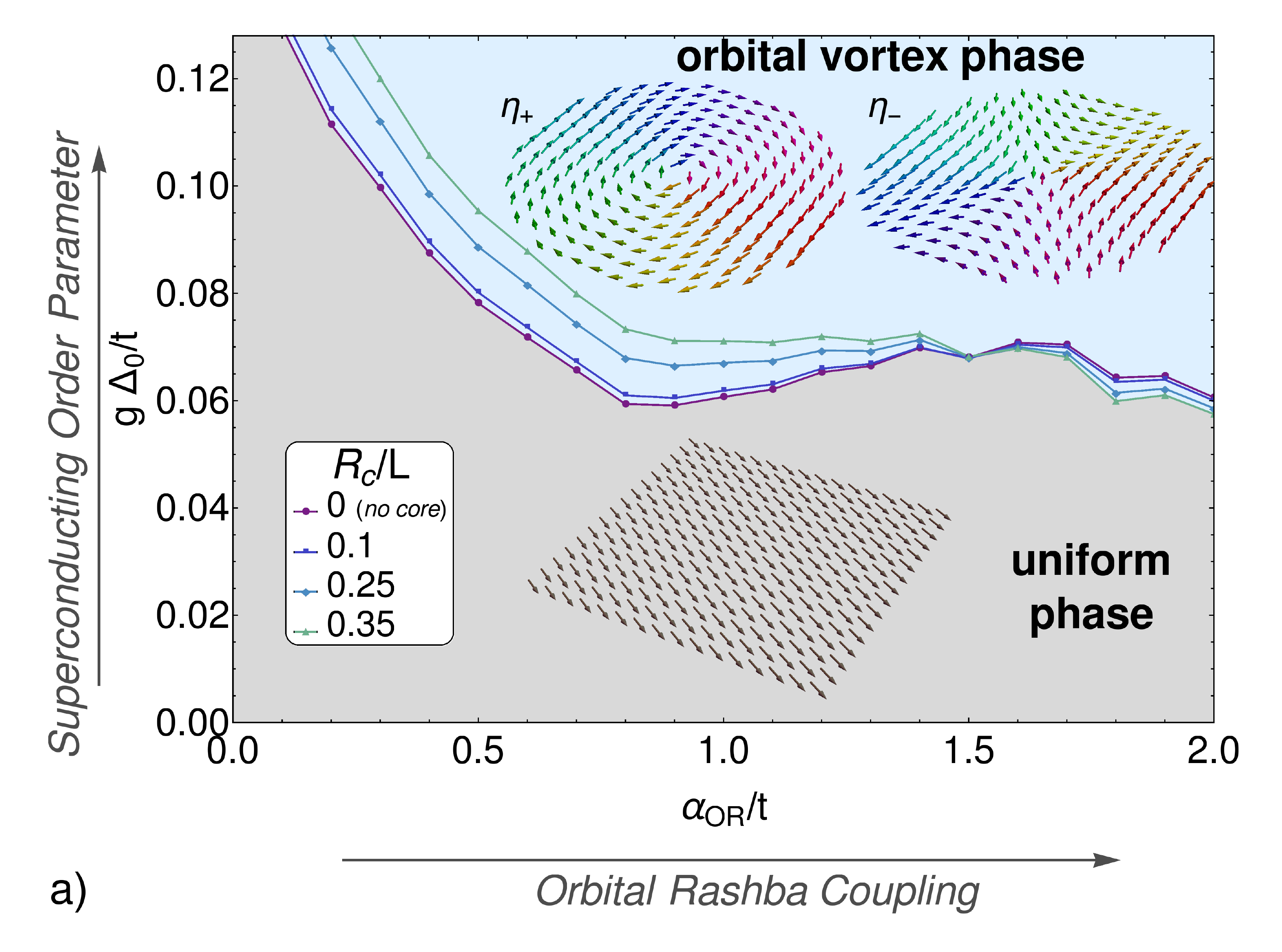}
 \includegraphics[width=0.49\textwidth]{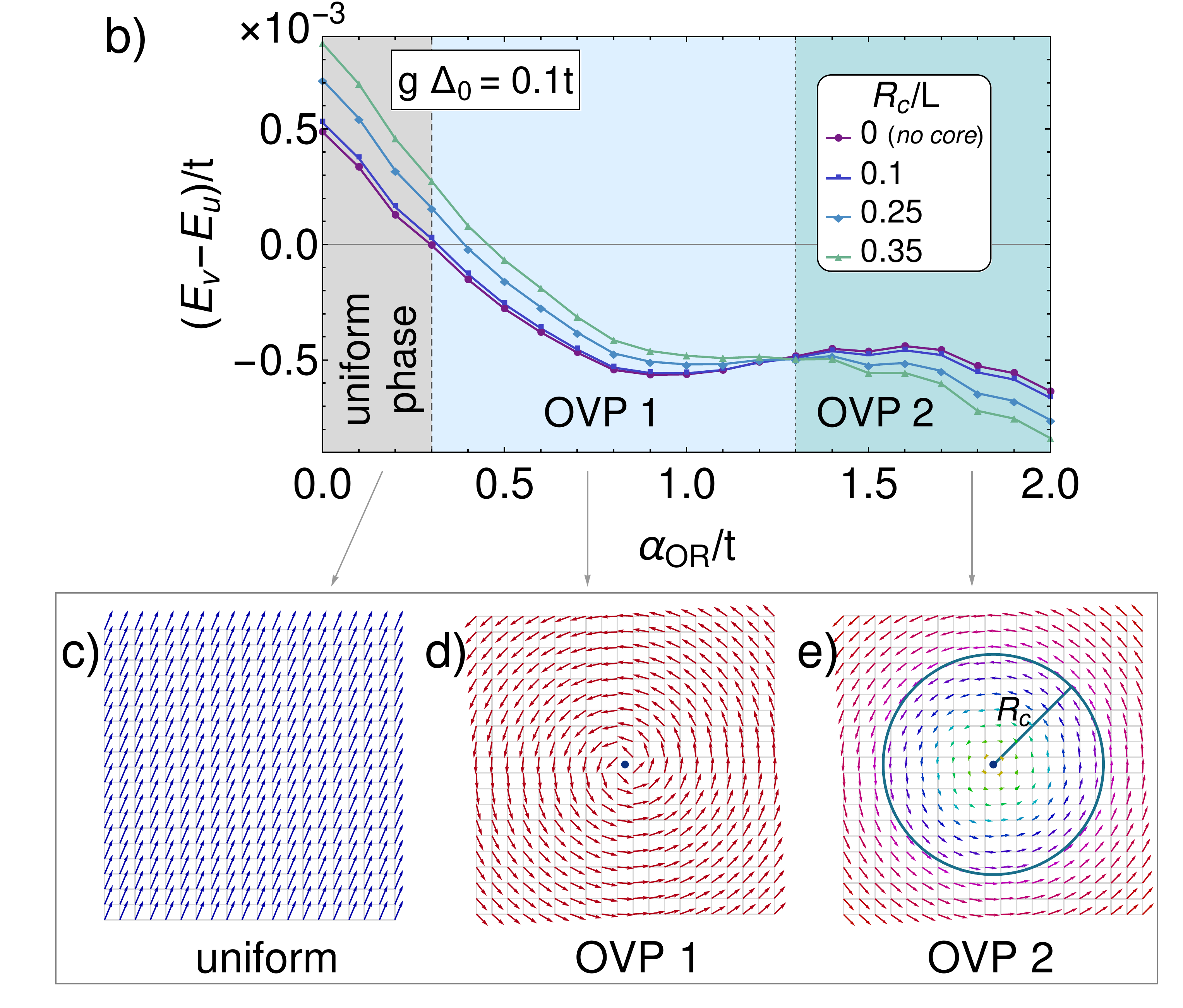}
\protect\caption{(a) Phase diagram with uniform and orbital vortex configurations as a function of the energy scale ($g\Delta_0$), setting the amplitudes of superconducting order parameters $\Delta_{\alpha\beta}$, and the orbital Rashba interaction ($\alpha_{\text{OR}}$). The purlple line ($R_c/L=0$) is the one presented in Fig.3(a) of main text. 
Here, the comparison among uniform and vortex states (besides the honogenous ($R_c=0$) case) includes also 
vortex with suppressed amplitudes ($R_c\neq 0$) of the superconducting order parameter within a distance $R_c$ from the core, with a spatial profile given by $f(r)=\tanh(r/R_c)$. The simulation refers to a system size $N_x \times N_y$ with $N_x=N_y=70$, while the model parameters are the same as in the main text.
(b) Energy difference between vortex ($E_v$) and uniform ($E_u$) phase as a function of $\alpha_{OR}$ for a representative amplitude of the pairing order parameter.
(c-e) Graphical representation of the orbital triplet order parameters ($\eta_{+,a}$) for the three possible regimes: (c) uniform, (d) homogeneous orbital vortex phase (called OVP1), and (e) orbital vortex phase (OVP2) with suppressed amplitudes of the superconducting order parameter within a distance $R_c$ from the vortex core. In (e) The length of the arrow is proportional to the amplitude of the superconducting order parameter. 
\label{f1}
}
\end{figure*}

\begin{figure*}[bt]
\includegraphics[width=0.98\textwidth]{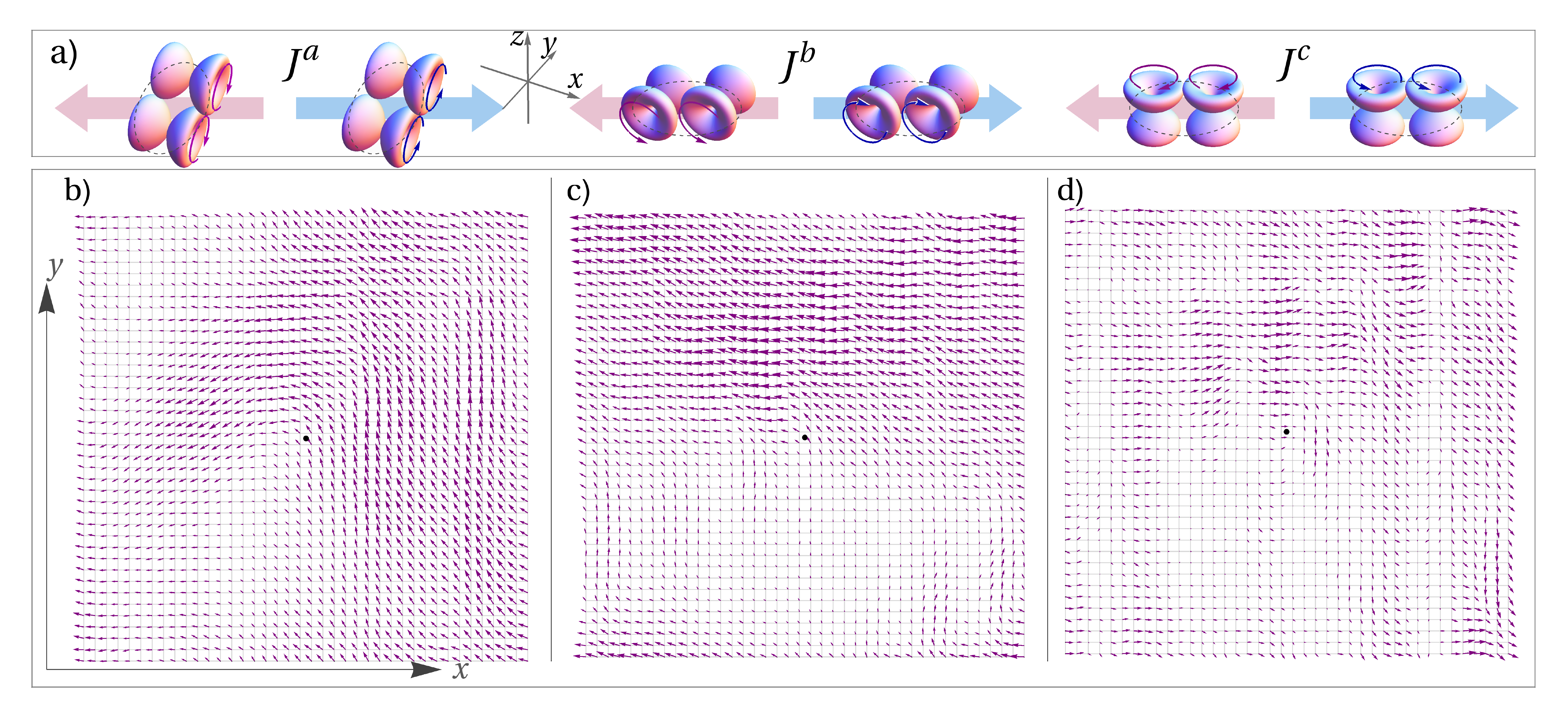} 
\protect\caption{(a) Illustration of electron pairs with parallel orbital moments having reversed sign of the orbital angular momentum projections and opposite velocity thus leading to a net orbital supercurrent without charge flow. (b-d) Spatial distribution of the orbital supercurrent [(b) $J^a$, (c) $J^b$, and (d) $J^c$] around the vortex core (black dot) for various atomic orbital orientations shown in (a). Note that $J^a$ is also shown in Fig2 of main text, we present it here again for a direct comparison with the other components.  The arrows indicate the direction of the orbital current on each bond, and the lengths its amplitude. 
The simulation refers to a system size $N_x \times N_y$ with $N_x=N_y=40$ while the following parameters of the 2D model Hamiltonian have been used: $\alpha_{\text{OR}}=1.0 t$,$\gamma=0.1 t$,$t_m=0.2 t$,$g\Delta_0=0.1 t$, $\mu=-0.4 t$ and for the vortex order parameters $\eta(r)$, $f_0=0.3$,$f_1=f_2=0.5$. 
\label{f2}}
\end{figure*}


\section{Symmetry of the superconducting order parameter} 
As it concerns the pairing function, we note that 
$\hat{\Delta}$ is
a $3\times 3$ matrix whose diagonal (off-diagonal) entries are given by the intra- (inter-) orbital spin-singlet order parameters. 
Indeed, the general structure of $\hat{\Delta}$ in the orbital space can be expressed in the following form, $\hat{\Delta}=\frac{1}{2} \sum_{l,m} g_{\alpha \beta} \Delta_{\alpha \beta} (\hat{L}_{\alpha} \cdot \hat{L}_{\beta}  +\hat{L}_{\beta} \cdot \hat{L}_{\alpha}$) with $\Delta_{\alpha \beta}(r)= \langle c_{r,0_\alpha \uparrow} c_{r,0_\beta\downarrow} \rangle$ the superconducting order parameters, while $g_{\alpha\beta}$ indicates the corresponding strength of the pairing interaction. 
Assuming the ${\mathcal C}_{4v}$ point group is not lowered by external perturbations only intra-orbital pairing amplitudes with $\Delta_{aa}=\Delta_{bb} \neq \Delta_{cc}$ are compatible with the point group symmetry as long as the $s$-wave spin-singlet channel is considered.
This is because the order parameter $\Delta_{aa} - \Delta_{bb}$ transforms as the one-dimensional irreducible representation (IRREP) $B_1$ of the ${\mathcal C}_{4v}$ point group and therefore can exist only in the $d$-wave pairing channel with the form factor $k_x^2-k_y^2$ since the latter also transforms according to the $B_1$ IRREP. Likewise, the order parameter $\Delta_{ab}$ transforms as the $B_2$ IRREP and therefore can appear again in the $d$-wave channel with the form factor $k_x k_y$. Note instead that the pairing order parameters $\Delta_{ac}$, $\Delta_{bc}$ cannot appear neither in the $s$-wave channel nor in the $d$-wave channel. Consider next a symmetry reduction to the  group ${\mathcal C}_{2v}$. In this case the order parameter component $\Delta_{ab}$ transforms according to the $A_2$ representation of ${\mathcal C}_{2v}$ and therefore can only appear in the $d$-wave channel with the form factor $k_x k_y$. On the contrary the order parameter component $\Delta_{aa} - \Delta_{bb}$ can also appear in the $s$-wave channel. As before, the order parameter components $\Delta_{ac}$, $\Delta_{bc}$ do not appear neither in the $s$-wave nor in the $d$-wave channel since they belong to the $B_{1,2}$ representations of ${\mathcal C}_{2v}$. Finally, an additional symmetry lowering to the group ${\mathcal C}_v$ does not change the situation. Hence the appearance of both the order parameter components $\Delta_{aa} - \Delta_{bb}$ and $\Delta_{ab}$ in the $s$-wave channel necessarily requires breaking of all rotation and vertical mirror symmetries. We point out, however, that when a single mirror symmetry is present the interorbital pairing $\Delta_{ac}$ that is left invariant under the mirror $M_y$ is allowed in the $s$-wave channel. This implies that a vortex state can be also obtained in the presence of mirror symmetries with the orbital moment of the electron pairs contributing to the vortex configuration lying in the mirror plane. 
In this context, it is also useful to comment on the role of the two-dimensional confinement and its relation with the amplitude of the $\Delta_{ac}$ and $\Delta_{bc}$ order parameters. The bands associated to the $a$ or $b$ and $c$ orbitals are subjected to a crystal field potential that splits their relative energies and, thus, would tend to suppress the $\Delta_{ac}$ inter-orbital pairing. However, this detrimental impact on $\Delta_{ac}$ or $\Delta_{c}$ is prevented by the orbital Rashba coupling which induces a mixing of the $(a,c)$ and $(b,c)$ orbitals such as to yield non-vanishing in-plane orbital moments at any given $k$ in the Brillouin zone.  

\begin{figure*}[bt]
\includegraphics[width=0.33\textwidth]{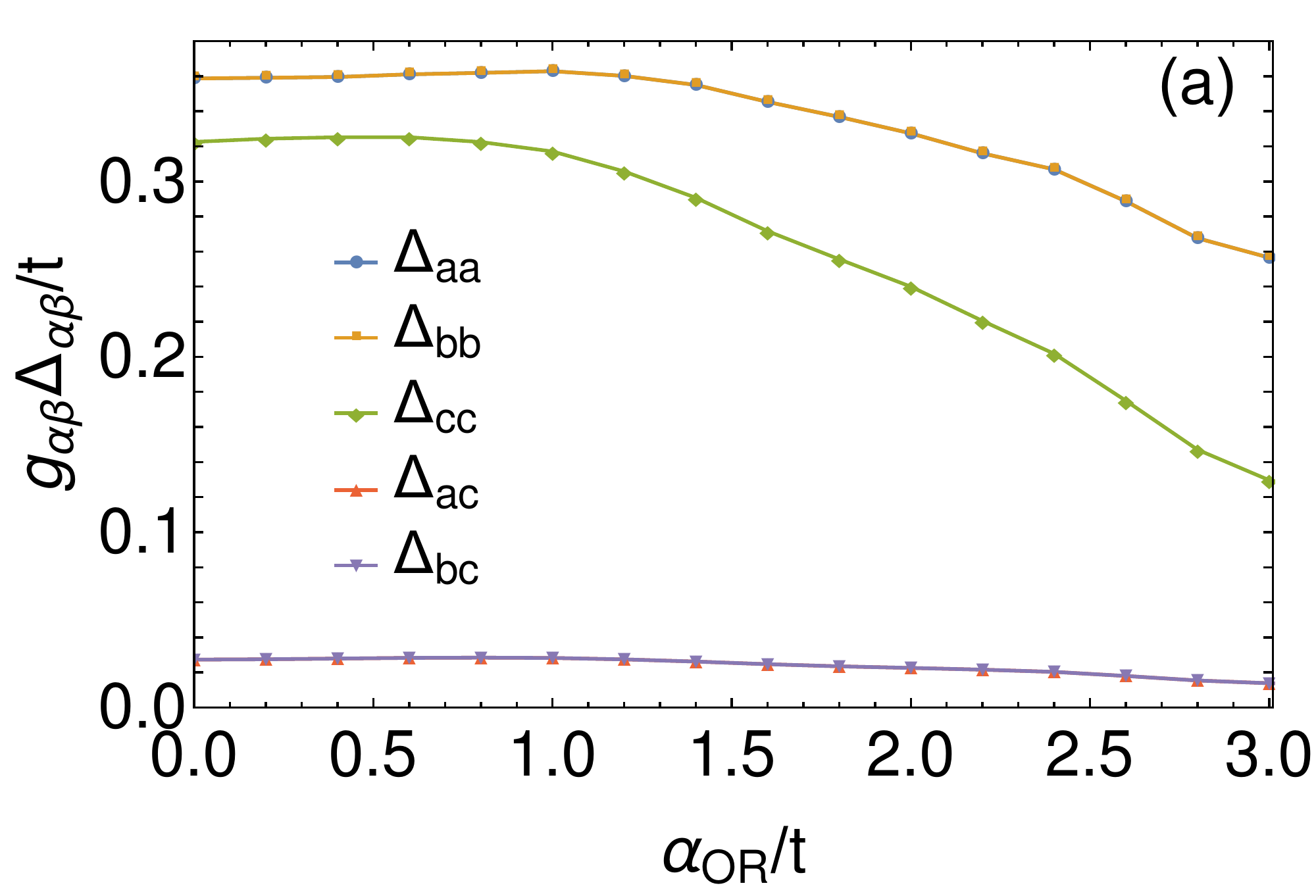}
\includegraphics[width=0.33\textwidth]{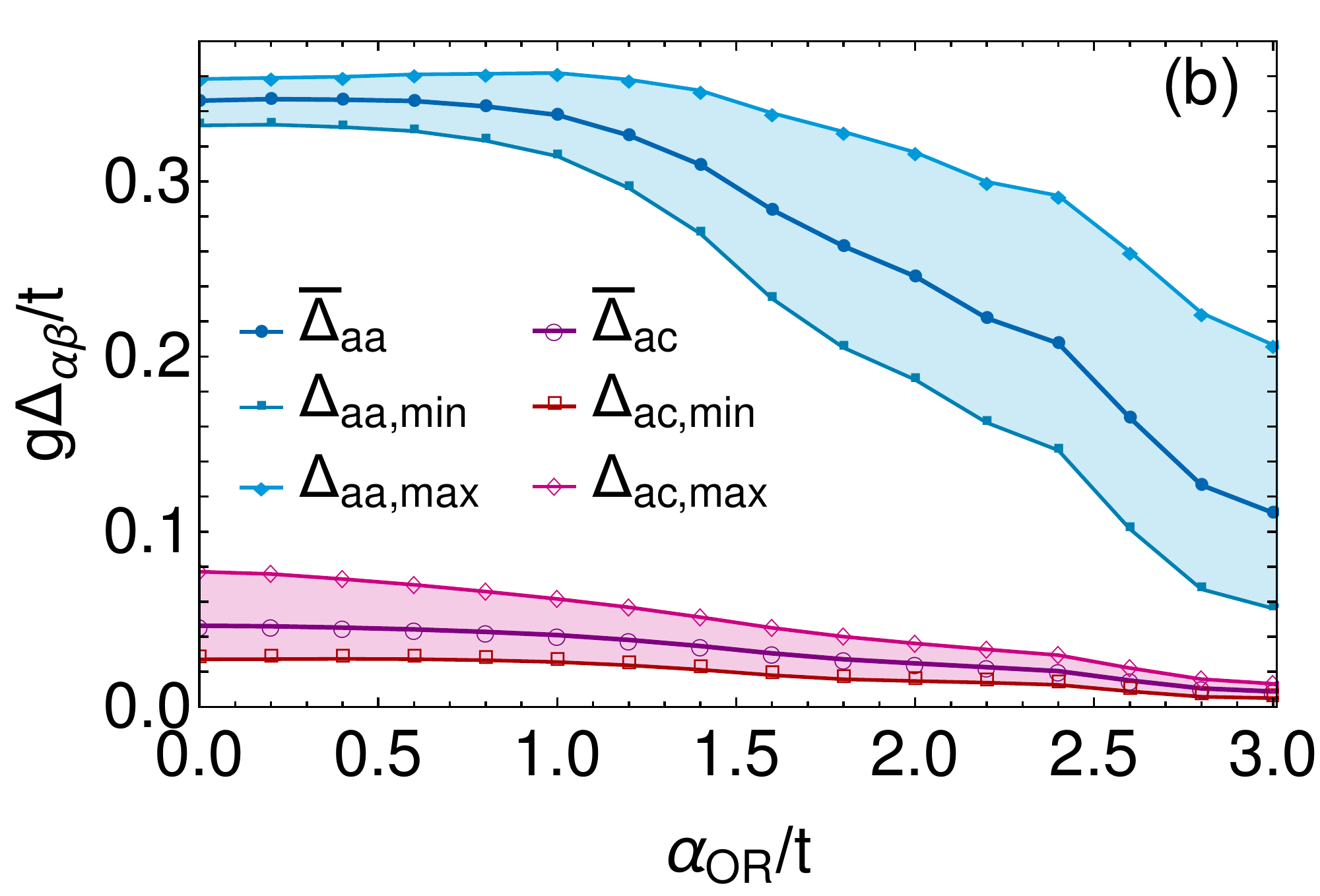}
\includegraphics[width=0.33\textwidth]{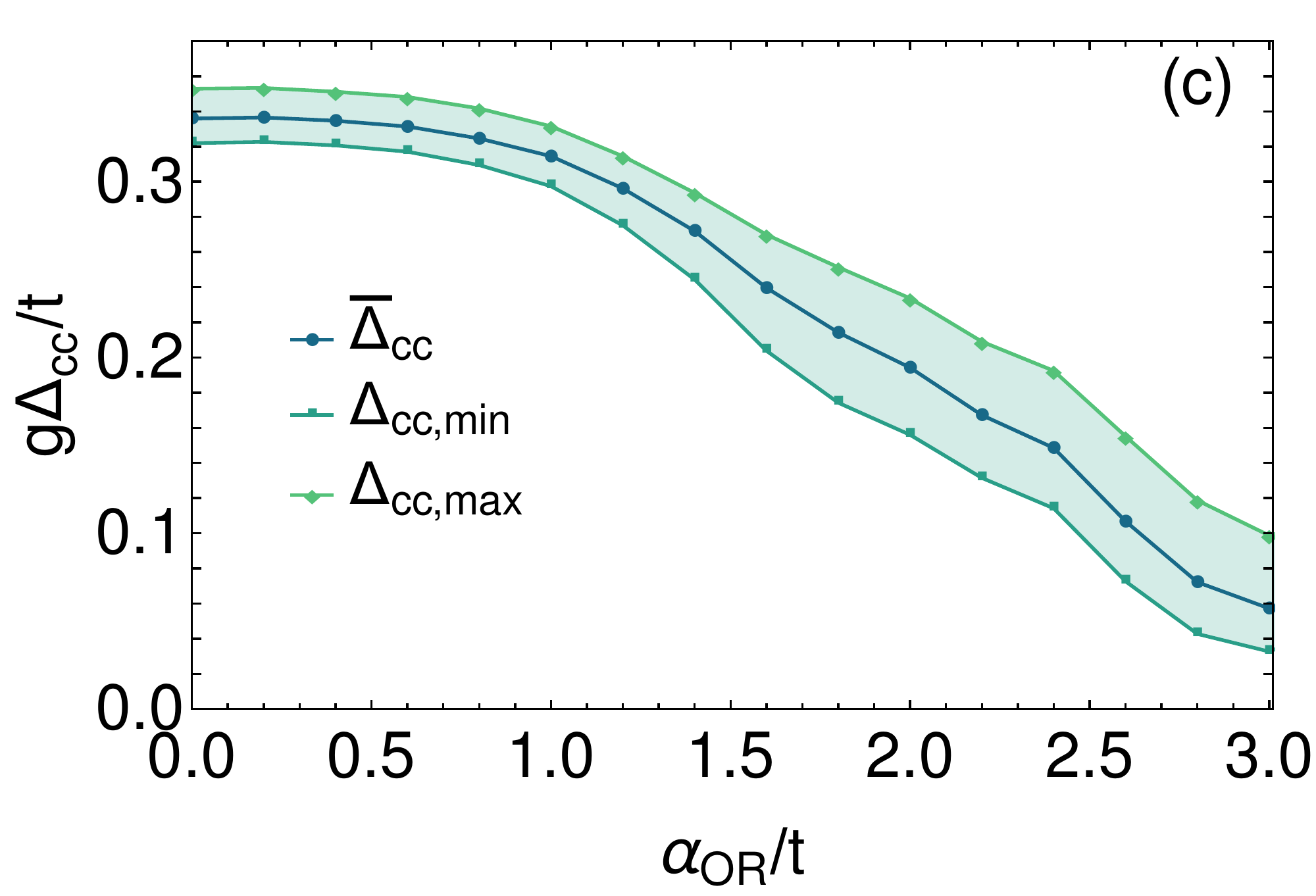}
\protect\caption{
Superconducting OP components $\Delta_{\alpha\beta}$  as a function of the orbital Rashba coupling 
$\alpha_{\text{OR}}$ obtained with self-consisten iterative procedure for a system of size $20\times 20$. In (a) we show the behavior 
$\Delta_{\alpha \beta}$ for the uniform solution, while in (b-c) that for the vortex state. In the latter situation the SC OP is not uniform over the lattice, hence we show the average values of $\Delta_{\alpha \beta}$ over all the sites ($\bar{\Delta}_{\alpha\beta}$) and the minimum and maximum values ($\Delta_{\alpha \beta,min}$ and $\Delta_{\alpha \beta,max}$). Specifically in (b) we present the solutions for $\Delta_{aa}$ and $\Delta_{ac}$, while $\Delta_{cc}$ is shown separately in (c) to avoid overlaps in the figure. 
\label{f3}
}
\end{figure*}

\begin{figure*}[bt]
\includegraphics[width=0.95\textwidth]{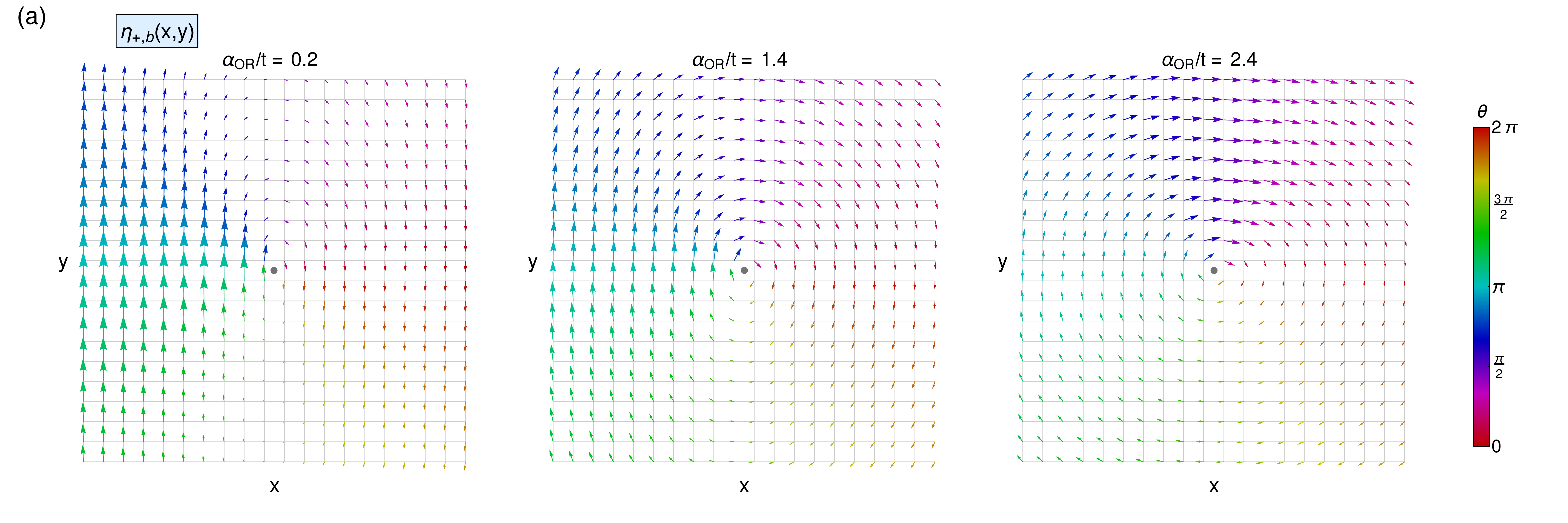}
\includegraphics[width=0.95\textwidth]{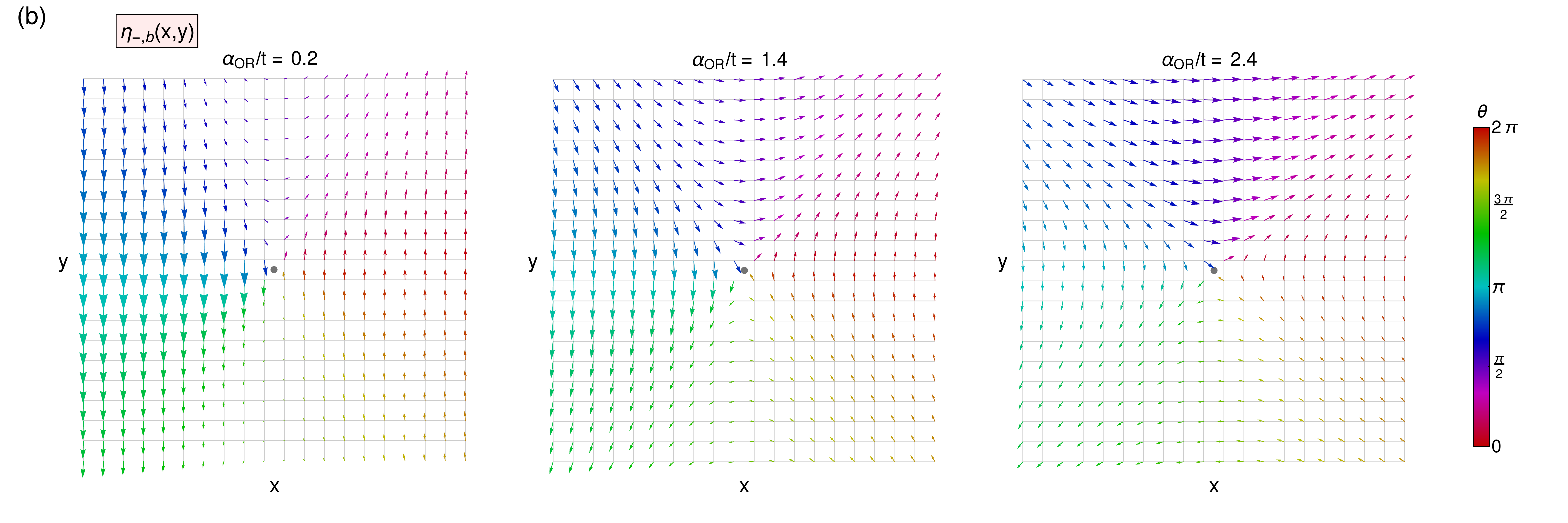}
\protect\caption{ (a-c) Graphical representation of the vortex order parameters $\eta_{+,b}$ and (d-f) $\eta_{-,b}$ with orbital moments oriented along $b$-axis, assuming that the vector components correspond to the real and imaginary parts of $\eta_\pm$, for three different values of the orbital Rashba coupling $\alpha_{\text{OR}}$. The solutions have been obtained with an iterative self-consistent procedure in real space for a system of size $(N_x=20)\times(N_y=20)$. The color gradient is used to emphasize  the winding of $\eta_\pm$. As an additional material two movies are attached which show the evolution of the vortex order parameter $\eta_{\pm,b}$ for several values of $\alpha_{\text{OR}} \in [0.2t, 3.0t]$. Similar patterns are also obtained for $\eta_{\pm,a}$, while $\eta_{\pm,c}$ is close to zero for the selected regime of parameters.
\label{f4}
}
\end{figure*}

\begin{figure}[bt]
\includegraphics[width=0.98\columnwidth]{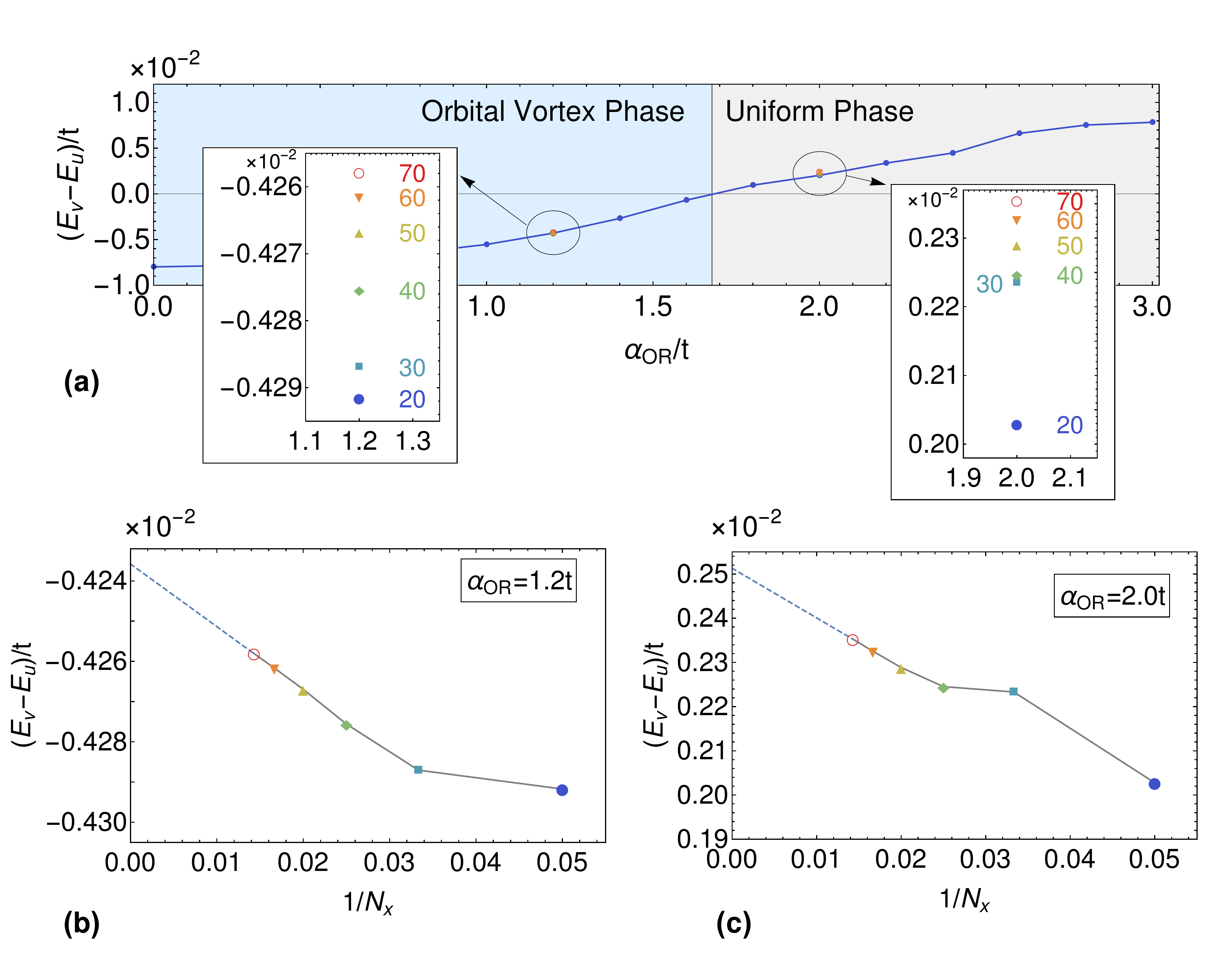}
\protect\caption{(a) Energy difference between the vortex solution ($E_v$) and the uniform one ($E_u$) as a function of the orbital Rashba coupling $\alpha_{\text{OR}}$, where the solutions have been obtained evaluating the superconducting order parameters 
$\Delta_{\alpha \beta}$ with an iterative self-consistent procedure in real space for a system of size $(N_x=20)\times(N_y=20)$ (already shown in the main text as Fig.3(b)). For two representative values of the orbital Rashba coupling, namely for $\alpha_{\text{OR}}=1.2t$ and $2.0t$, we solved the problem by increasing the lattice size dimension up to $N_x=N_y=70$ (see insets for a zoom on the two regions). The numbers in the insets refer to the number of sites on each side of the square lattice ($N_x=N_y$). (b)-(c) Plots of the energy difference ($E_v-E_u$) as a function of $1/N_x$ for the two values of $\alpha_{\text{OR}}$ considered (written in the figures). The dashed lines are the extrapolation of the straight lines of the last three points. The extrapolation to the thermodynamic limit indicates a linear behavior for the 
energy difference. The limiting value confirms the trend obtained for the finite size calculation.
\label{f5}
}
\end{figure}

\section{Vortex phase} 

\subsection{Vortices with uniform or radially inhomogeneous amplitudes}

Here, we consider the competition between the uniform phase and the vortex state by taking an ansatz for the amplitude of the superconducting order parameters. 
We recall that $\eta_a$, $\eta_b$ and $\eta_c$ are expressed as:
\begin{eqnarray}
\eta_{\pm,a}(r)&&=\left[ \frac{1}{2} \left(-\Delta_{bb}(r) + \Delta_{cc}(r)\right) \mp i \Delta_{bc}(r) \right]\,, \\
\eta_{\pm,b}(r)&&=\left[ \frac{1}{2} \left(-\Delta_{aa}(r) + \Delta_{cc}(r)\right) \pm i \Delta_{ac}(r) \right]\,, \\
\eta_{\pm,c}(r)&&=\left[ \frac{1}{2} \left(-\Delta_{aa}(r) + \Delta_{bb}(r)\right) \pm i \Delta_{ab}(r) \right]\,,
\end{eqnarray}

We firstly consider a solution for the $\eta(r)=f(r) \exp[i \theta(r)]$ order parameters that can have a uniform amplitude $f(r)$ or with a spatial profile that is given by $f(r)=\tanh(r/R_c)$, thus suppressed within a distance $R_c$ from the core.
In Fig. \ref{f1}(a) we show that as a function of the orbital Rashba coupling one can induce a transition from a uniform (Fig. \ref{f1}(b)) configuration to different types of vortex states with homogeneous (Fig. \ref{f1}(c)) or suppressed amplitudes (Fig. \ref{f1}(d)) of the superconducting order parameter within a distance $R_c$ from the core, with a spatial profile that is given by $f(r)=\tanh(r/R_c)$.
The behavior of the energy difference is explicitly shown in Fig. \ref{f1}(a) for a given amplitude of the pairing interaction (i.e. $g \Delta_0=0.1 t$).
We find that at $\alpha_{\text{OR}}\sim 0.3 t$ a transition occurs between the uniform phase and a vortex configuration with constant amplitude of the order parameters (i.e. $R_c=0$). 
On the contrary,
at larger values of the orbital Rashba coupling ($\alpha_{\text{OR}} > 1.3 t$) a vortex state with a normal core radius $R_c$ that is comparable to the system size $L$ becomes energetically more favorable than the configuration with vanishing $R_c$. The computation refers to a system size $N_x \times N_y$, with $N_x=N_y=70$, while the following parameters of the 2D model Hamiltonian have been used: $\alpha_{\text{OR}}=1.0 t$,$\gamma=0.1 t$,$t_m=0.2 t$,$g\Delta_0=0.1 t$, $\mu=-0.4 t$, with $f_0=0.3$ and $f_1=f_2=0.5$ for the superconducting order parameters.

For this type of vortices, we have an orbital supercurrent with a characteristic winding of the angular momentum flow. The vorticity is regular and of the same type for $J^a$ and $J^b$ while it has a sort of more turbulent aspect for $J^c$ as shown in Fig. \ref{f2}. 
The circulation of electron pairs with a three-dimensional orientations of the orbital moments are fingerprints of the orbital vortex. 

\subsection{Vortex state: self consistent solutions, phase diagram and size dependence} 
Here, we provide details about the evolution of the self-consistent solutions for the intra- and inter-orbital order parameters as a function of the orbital Rashba coupling (Figs. \ref{f3},\ref{f4}). 
Concerning the evaluation of the superconducting order parameters at a given site $r$, $\Delta_{\alpha\beta}(r)=\frac{1}{N_x N_y} \sum_{n} g_{\alpha\beta}\,\langle n| c_{r,0_\alpha\uparrow}(r) c_{r,0_\beta\downarrow}(r) |n \rangle$, with $(\alpha,\beta)$ being the orbital indices, we performed it by self-consistently computing the trace of the pairing operator for the spin-singlet channel, $\hat{P}^{\alpha\beta}(r) =c_{r,0_\alpha\uparrow}(r) c_{r,0_\beta\downarrow}(r)$, over all the eigenstates $|n\rangle$ of the Hamiltonian associated to negative energies $E_{n}<0$ at zero temperature (at finite temperature the trace is over all energy configurations weighted by the Fermi function). Since the eigenstates $|n\rangle$ depend on $\Delta_{\alpha\beta}(r)$ and the orbital Rashba interaction couples the crystal wave vector with the in-plane orbital moments, the gap equations of the orbital dependent order parameters are strongly coupled between each other. The procedure is iterated by evaluating the order parameters for all the sites in the cluster upon reaching the desired accuracy.

More specifically, the trend is obtained for a system with size $N_x=N_y=20$ considering the uniform (Fig. \ref{f3}a) and the vortex solutions (Figs. \ref{f3}b,c). For the vortex solutions we have a distribution of amplitudes and we report only the average ($\bar{\Delta}$) together with minimum ($\Delta_{\text{min}})$ and maximum value ($\Delta_{\text{max}}$). The overall behavior indicates that the effect of the orbital Rashba coupling is to reduce the amplitude of the orbital dependent order parameters. The vortex pattern in self-consistent is presented in Fig. \ref{f4} for several values of the orbital Rashba coupling.

Finally, we determine the size dependence of the energetics of the vortex solution as evaluated within the self-consistent procedure. As shown in Fig. \ref{f5}, the computation for different sizes of the system with $N_x=N_y=L$ and $L=20,30,40,50,60,70$ demonstrates that the stability of the vortex state is not affected by size variation.
